\documentclass[12pt,preprint]{aastex}

\newcommand{\etal}{{et al.}\ }
\newcommand{\eg}{{e.g.,}\ }

\shorttitle{White Dwarfs in Open Clusters}
\shortauthors{von Hippel}

\begin{document}

\title{From Young and Hot to Old and Cold: Comparing White Dwarf Cooling
Theory to Main Sequence Stellar Evolution in Open Clusters}

\author{Ted von Hippel}
\affil{The University of Texas at Austin, Department of Astronomy, 
1 University Station C1400, Austin, TX 78712-0259, email: ted@astro.as.utexas.edu}

\begin{abstract}
I explore the current ability of both white dwarf cooling theory
and main sequence stellar evolution theory to accurately determine
stellar population ages by comparing ages derived using both techniques
for open clusters ranging from 0.1 to 4 Gyr.  I find good agreement
between white dwarf and main sequence evolutionary ages over the entire
age range currently available for study.  I also find that directly
comparing main sequence turn-off ages to white dwarf ages is only weakly
sensitive to realistic levels of errors in cluster distance, metallicity,
and reddening.  Additional detailed comparisons between white dwarf and
main sequence ages have tremendous potential to refine and calibrate both
of these important clocks, and I present new simulations of promising open
cluster targets.  The most demanding requirement for these white dwarf
studies are very deep ($V \geq$ 25--28) cluster observations made
necessary by the faintness of the oldest white dwarfs.  
\end{abstract}


\section{Introduction}

White dwarf cooling theory currently provides the most reliable age for
the Galactic disk (Winget \etal 1987; Oswalt \etal 1996; Leggett, Ruiz, \&
Bergeron 1998; Knox, Hawkins, \& Hambly 1999), whereas main sequence
stellar evolution provides the most reliable age for the Galactic halo
(e.g., Salaris \& Weiss 2002; Krauss \& Chaboyer 2003).  In order to
understand the detailed sequence of formation of the Galactic disk and
halo, as well as the thick disk, bulge, and local satellite galaxies,
these two time scales need to be placed on the same absolute age system.
The only current empirical approach available to inter-calibrate these two
age systems is to derive white dwarf (WD) cooling ages and main sequence
turn-off (MSTO) ages for a number of single-age stellar populations over a
wide range of ages.  Since old WDs are faint ($M_{\rm V} >$ 16), a further
constraint is that the stellar population is within a few kiloparsecs, or
the target objects become too faint ($V~\geq$~30) to observe.  In
addition, each stellar population should have a single heavy element
abundance, yet the total sample of stellar populations should cover a
range of heavy element abundances, so that detailed studies can search for
any metallicity effects on age potentially missing from either theory.
Many of the most well-known open clusters fit these needs for nearby,
single-age, single-metallicity stellar populations well, and furthermore a
sample of some of the most favorable open clusters cover a wide range of
ages and a substantial range of heavy element abundances.  Globular
clusters can be used to extend such a study to even greater ages and lower
metallicities, though at present only a few globular clusters are near
enough for observations to be performed to the limit of their coolest
WDs.

The first studies to explicitly compare WD cooling and MSTO ages for star
clusters were those of Claver (1995) and von Hippel, Gilmore, \& Jones
(1995).  These studies demonstrated that the WD sequence of a cluster
shows a low-luminosity terminus that is determined by the cooling age
of the WD population and its comparatively short pre-WD evolution.
Subsequent studies (Richer \etal 1998; von Hippel \& Gilmore 2000; von
Hippel 2001; Claver \etal 2001) have shown that for open clusters, WD
cosmochronology and main sequence stellar evolution give similar cluster
ages.  WD age studies have now been extended to one globular cluster
(NGC 6121 = M4) where a WD age has been derived (Hansen \etal 2002),
disputed (De Marchi \etal 2003), and defended (Richer \etal 2004).
Regardless of the observational reliability of the M4 study, WD cooling
models are not yet at the point where they can give reliable values for
the great ages of globular clusters (Hansen \& Liebert 2003), due to
uncertainties in the theory of cool WDs.  The pay-off of an independent,
accurate, and precise age determination for globular clusters via the WD
cooling technique is enormous, however, and so its calibration is of
fundamental importance.  Such an independent age determination would
either support or contradict ages derived from main sequence stellar
evolution and should allow a more precise comparison between the age of
the Galaxy and the now precise, if model-dependent, age for the Universe
from WMAP observations (Bennett \etal 2003).

Additionally, with the Advanced Camera for Surveys (ACS) in the Hubble
Space Telescope (HST), it is easy to reach the terminus of the white dwarf
sequence in a number of open clusters and possible to reach the terminus
in a few of the nearest globular clusters.  Cluster observations of
sufficient depth, $V > 26$, are often not possible with the current
generation of 8-10m telescopes, however, not only because of the low flux
levels, but also because of the many contaminating, compact background
galaxies with the approximate colors of cool white dwarfs.  For those
clusters where the faintest WDs are brighter than the limit of 8-10m
observing capabilities, it may be possible with proper motions derived
from second epoch observations obtained some years later, to remove the
contaminating background galaxies.  Next generation 20- to 30-meter
ground-based telescopes should also be able to make these very deep
observations, especially once their adaptive optics systems are pushed
blueward into the $I$-band, where contaminating faint background galaxies
can be spatially resolved.  These current and forthcoming instrumental
capabilities, along with recent improvements in both stellar evolution and
WD cooling theory, justify a renewed look at the current state-of-the art
in both theories and an examination of which clusters would be best suited
to the improved techniques and instrumentation.

\section{White Dwarf Versus Main Sequence Ages in Open Clusters}

How do the ages derived by main sequence stellar evolution and WD cooling
ages compare in open clusters?  Table 1 presents a list of open clusters
for which a WD (cooling plus precursor) age has been derived, along with
recent age determinations from main sequence stellar evolution studies.
In choosing which MSTO ages to incorporate, I have used those studies that
relied on models with core convective overshoot as past studies of NGC
2420 (von Hippel \& Gilmore 2000) and M 37 (Kalirai \etal 2001) have found
a better match between core convective MSTO ages and WD ages for these
clusters.  I do not include the new WD cooling results for M4 in Table 1
due to the current extrapolation in WD theory necessary to date this
cluster.  The first column lists the cluster name, the second column lists
the derived WD age and $\pm1 \sigma$ age errors, the third column lists
the main sequence turn-off age and $\pm1 \sigma$ age range from one or
more recent studies, and the fourth column points to a reference list for
the WD and MSTO ages.  Prior discussions of the comparison of WD and MSTO
ages in open clusters can be found in von Hippel (2001) and Hansen
(2004).

To graphically present the main results of Table 1, Figure 1 presents MSTO
versus WD ages for these clusters.  The WD age for the Hyades is a lower
limit since 50 to 90\% of the Hyades has likely evaporated (Weidemann
\etal 1992), possibly taking with it some of the oldest WDs.  (Strictly
speaking, cluster WD ages, unlike field star white dwarf ages, always
provide a lower age limit since the oldest WDs could be missing due to
partial cluster evaporation.)  The WD age for the oldest cluster plotted
here, M 67, is based on a statistically measured WD luminosity function
for the cluster after subtraction of a comparison field (Richer \etal
1998).  For this cluster, candidate cluster WDs have not yet been
spectroscopically confirmed, and therefore the WD versus MSTO comparison
at this age is not yet firmly established.  Figure 1 demonstrates that
there is a good overall agreement between cluster ages derived via the two
different techniques.  Assuming one uses modern overshoot ages, the WD and
MSTO ages agree for six open clusters within their age errors, and the
remaining cluster (the Hyades) is consistent with this age agreement.
Stellar evolution and white dwarf cooling provide consistent ages over the
broad age range that we can currently test, 0.1 to at least 2 Gyr and
probably to 4 Gyr.

To further study the utility of Figure 1, I calculated the effects of
typical observational errors in this diagram.  I considered three types of
observational error.  For the first type of error, I over-estimate the
distance by 0.2 mag, due for example to a combination of photometric
calibration errors and errors in deriving the main sequence fitting
distance.  For the second type of error, I over-estimate the metallicity
by 0.2 dex.  For the third type of error, I also over-estimate the
metallicity by 0.2 dex, but now I also adjust the reddening to compensate
for the color change in the main sequence turn-off caused by the
metallicity error.  One or more of these three types of error are present
in just about any study of open or globular clusters, and the values have
been set to be in the range of typical to somewhat conservative.  To
convert these assumed errors into an age error I use the cluster models
outlined in the next section.

Figure 2 presents the effect of these three types of error on the derived
MSTO age versus the ratio of the WD-to-MSTO age over the range of 100 Myr
to 4 Gyr.  The thick lines show the effect of over-estimating distances by
0.2 mag, which forces the cluster MSTO and WD ages to be under-estimated
since both turn-off stars and WDs are then assumed to be brighter than
they actually are.  The thin lines show the effect of over-estimating
metallicity by 0.2 dex, in which case overly red isochrones are force-fit
to the turn-off stars and the MSTO ages change.  In the cluster models
there is no significant change in the WD ages with this small metallicity
change, as metallicity is assumed to enter only through the
slightly-modified ages of the high mass precursors to the oldest WDs, and
this latter effect is included.  The dotted lines show the effect of
over-estimating metallicity by 0.2 dex, then compensating by
under-estimating reddening to keep the cluster turn-off at the same
color.  This type of error affects the WD ages since a change in reddening
changes the apparent magnitude of the WD terminus.  Of course, for some
real clusters with low reddening, it is not possible to lower the
reddening enough to match some erroneous metallicity determinations, and
in these cases there are additional limitations on errors in this third
category.  It is comforting to see that the net effect of realistic errors
on the WD and MSTO ages are small, typically changing the ratio of ages by
$\sim$10\% through much of this age range, though reaching a maximum for
distance errors amounting to $\sim$50\% at 4 Gyr.  The derived absolute
MSTO ages change by $<$ 25\%, except at 4 Gyr where the effect of an
overestimated distance by 0.2 mag lowers the implied MSTO age by 40\% to
$\sim$2.4 Gyr.  In this case the implied WD age drops by 10\% to $\sim$3.6
Gyr.

Figure 2 also shows interesting responses to these sources of error for
clusters as a function of age.  While over-estimating distances always
causes an under-estimate in both the MSTO and WD ages, the ratio of these
two under-estimated ages changes around 2 Gyr.  For younger clusters, the
sensitivity of the location of the WD terminus to age is slightly greater
(up to 10\%) than the sensitivity of the location of the MSTO to age, and
thus the WD approach under-estimates age to a slightly greater degree than
the MSTO approach.  At 2 Gyr, both the MSTO and WD age techniques are
equally sensitive to direct displacements in luminosity, and at 4 Gyr the
MSTO age is the more sensitive technique, and thus under-estimates age to
a greater degree than the WD technique.  The change in relative
sensitivity of the two techniques to metallicity errors near 1 Gyr is
predominantly due to the effect of metallicity changes on the MSTO, since
the WD technique is largely insensitive to metallicity errors.  The
dominant effect is that the MSTO becomes fainter if metallicity is
over-estimated for clusters between 0.1 and 0.4 Gyr, whereas the MSTO
becomes brighter for clusters between 1 and 4 Gyr, at least for our input
stellar models (Girardi \etal 2000) near solar metallicity.  The final
type of error, with combined and offsetting errors in metallicity and
reddening, is somewhat more complicated.  For the youngest clusters and
our input models, a change in metallicity causes only a small color
change, and therefore induces only a small error in reddening.  The color
change increases rapidly as age increases to 1 Gyr, reaching a maximum
color difference of $B-V \approx 0.08$ mag for a change in metallicity of
0.2 dex, then decreases slowly as age increases to 4 Gyr.  The basic
effect is that the MSTO ages are over-estimated with this particular
coupling of errors (over-estimating metallicity by 0.2 dex, then
compensating by under-estimating reddening), since the under-estimated
reddening forces one to compensate and assume the MSTO is fainter than it
really is.  The effect on the implied WD ages from this type of error is
determined by the size of the reddening error and the sensitivity of the
WD technique to shifts in luminosity as a function of age.  For young
clusters the effect is small since the under-estimated reddening is
small.  As the offsetting error in reddening increases, its effect on the
WD ages becomes significant.

In Figure 3, I again present the three categories of error studied in
Figure 2, now in the same observational plane of Figure 1.  In this
diagram of direct age comparison, it is clear that the effects of typical
errors are to move the derived ages largely along the one-to-one
correspondence line.  This is both good news and bad news for comparing WD
and MSTO ages in clusters.  The bad news is that independent ages via the
two techniques offer little leverage on the other cluster parameters of
distance, metallicity, and reddening.  The good news is that the relative
ages of the clusters change little with these types of errors, especially
for clusters younger than 2 Gyr, and so comparing these two ages remains a
powerful way of checking on the consistency between main sequence
evolutionary theory and white dwarf cooling theory.  At 4 Gyrs the
departure from the correspondence line is greater for a distance error of
0.2 mag, but fortunately the cluster in that position in Figure 1, M 67,
is one of the best studied old open clusters, and its distance uncertainty
is likely to be substantially less (Sarajedini \etal 2004, and references
therein).  Figure 3 also demonstrates at least part of the reason why the
clusters in Figure 1 agree so well in their MSTO and WD ages - this
diagram, at least in this age range, is insensitive to reasonable errors
in cluster distance, metallicity, or reddening.

While the agreement between MSTO and WD ages should give us confidence in
both methods of age dating stellar populations with ages of $\leq$ 4 Gyr,
we need to remain cautious when interpreting and comparing ages for older
populations such as the Galactic disk and halo.  It is also important to
increase the precision of age dating clusters younger than 4 Gyr, as
increased precision could help tease out subtle effects that may not be
correctly modeled in white dwarf or main sequence stellar evolution.  Such
effects could include the degree of core overshooting and its metallicity
dependence as well as the transition from the CNO bi-cycle to PP burning
in main sequence stars, or for white dwarfs, mass loss on the asymptotic
giant branch and the initial-final mass relation, envelope effects and
dredge-up, and carbon-oxygen phase separation during crystallization.
Future observations and analyses of more star clusters over a broad age
range are therefore needed to increase age dating precision, refine
current WD and main sequence theory, and to improve the analyses of ages
for older stellar populations, particularly those older than 4 Gyr.
Additionally, age studies for clusters nominally at the same age, but with
different metallicities, are needed to test the dependence of both WD
cooling theory and main sequence theory on heavy element abundance, since
we know the most about the high metallicity Sun and solar neighborhood yet
wish most to age date the low metallicity Galactic halo.

\section{Improving White Dwarf Age Determinations: New Observations and
New Techniques}

How do we build upon and refine the present, carefully collected set of
observations and results comparing WD ages and MSTO ages?  Certainly, HST
with the ACS offers new capability, and capability well-matched to this
problem.  The next generation of very large ground-based telescopes should
also easily recover the coolest white dwarfs in many open clusters and
probably also in a few of the nearest globular clusters, especially if
their adaptive optics systems can be pushed into the $I$-band, a
wavelength sensitive to cool WDs and to background galaxy morphology.  To
motivate further studies comparing MSTO to WD ages in clusters, I present
here a handful of simulated clusters that appear to be good candidates for
investigation.  These simulations allowed me to explore the trade-offs
between cluster parameters, the contaminating Galactic field, and
observational difficulty, and thereby reject some clusters that would be
poor candidates for WD age studies.

In Figures 4-10, I present simulated $VI$ color magnitude diagrams (CMDs)
for the open clusters NGC 1245, NGC 2204, NGC 2243, NGC 2360, NGC 2506,
NGC 2660, and NGC 7789.  The major characteristics of these clusters are
listed in Table 2.  These clusters were chosen since they are relatively
nearby, rich in members, have moderate or low inter-stellar absorption,
and in the important age range for WD age studies -- these are good
candidates for HST/ACS observations.  Two older clusters, NGC 188 ($\sim7$
Gyr; Sarajedini \etal 1999) and NGC 6791 ($\sim8$ Gyr; Chaboyer \etal
1999) are not presented here, though they are good, but difficult,
observational targets.  At this point, I am not simulating such clusters
since their highest mass, crystallizing WDs are not yet modeled (see
below) in a sufficiently realistic manner for ages greater than 5 Gyr.

The cluster simulations of Figures 4-10 incorporate a Miller \& Scalo
(1979) initial mass function (IMF), main sequence and giant branch stellar
evolution time scales of Girardi \etal (2000), the initial (main sequence)
- final (white dwarf) mass relation of Weidemann (2000), WD cooling time
scales of Wood (1992), and WD atmosphere colors of Bergeron \etal (1995).
Each star is randomly drawn from the IMF and, based on an appropriate
binary star fraction (here set to 50\%, typical for open clusters),
randomly assigned to be a single star or a binary with a companion also
randomly drawn from the IMF\footnote{The implied age from either the MSTO
technique or the WD technique is insensitive to the IMF.  The IMF serves
only to populate the particular mass region that is currently at the MSTO
or at the faint end of the WD sequence.  If there are insufficient stars,
particularly if the cluster is young, then the few cluster stars coupled
with the IMF can create an additional, statistical uncertainty to locating
the MSTO or faintest WDs.  Binaries of nearly any mass ratio have a
similar effect.  WDs in such systems are generally not recognized and MSTO
stars in such systems are found brighter and generally redder than the
MSTO, and therefore they do not help define the MSTO.}.  Other stellar
evolution (\eg Yi \etal 2001; Baraffe \etal 1998; Siess, Dufour, \&
Forestini 2000) and WD cooling (\eg Benvenuto \& Althaus 1999; Hansen
1999) models could have been used, but for the present purposes, these
often-used models adequately cover parameter space.  The simulated color
magnitude diagrams incorporate realistic photometric errors, for
observational depths set to match $V = 26$ or 0.5 magnitudes beyond the WD
terminus, whichever is fainter, at S/N = 15\footnote{From experience,
S/N=15 is required to obtain good morphological rejection of background
galaxies at HST resolution.  By placing this value 0.5 mag below the
expected terminus of the WD sequence, one has a bit of insurance against
the cluster being older than expected.  While not strictly necessary, even
if the cluster is as old as expected, the clear gap below the WD sequence,
now devoid of contaminating background galaxies, makes a convincing case
that the WD terminus has been properly identified.  If the WD terminus is
at $V < 25.5$, it is easiest to still observe to $V = 26$, since this
depth can be achieved in a single HST orbit.}.  The simulated CMDs also
incorporate field stars as predicted by the model of Reid \& Majewski
(1993).  These simulations do not include mass segregation or other
dynamical processes, which can be important in open clusters, especially
for the lowest mass stars, but which typically have little effect on the
measured WD mass fraction (von Hippel 1998; see also Hurley \& Shara 2003
who find that the WDLF and mass function are insensitive to dynamical
effects at 0.5 to 1 half-mass radii).  The cluster stars are presented as
filled circles, whereas the field stars are presented as 6-pointed
symbols.  One sigma error bars are included for the cluster white dwarfs
only.  To guide the eye to the expected location of the WDs in these CMDs,
the cooling track for a 0.7 $M_\sun$ WD cooler than $T_{\rm eff}$ = 15,000
K from Hansen (1999) is presented in each CMD.  A clear limitation of
these simulations is that stars with masses $\leq 0.15 M_\sun$ are not
included, thus the unrealistic lower limit to the main sequence and the
limited variety of binaries among the lowest mass stars.  Since the focus
of this study is on stars that can become WDs, this simplification is
merely one of presentation.  The number of simulated cluster stars should
be approximately the number that one would observe near the cluster
centers in an HST/ACS field (3.37 $\times$ 3.37 arc minutes).  This number
is determined by normalizing the star counts of each cluster to the known
brighter stars in these clusters from the published CMDs (see reference
list in Table 2).  This technique, though dependent on an extrapolation of
the IMF, has worked well enough in the past, providing the approximate
predicted number of cluster stars in HST/WFPC2 observations of NGC 2420
(von Hippel \& Gilmore 2000).

Even a quick study of these simulated clusters shows some of the
difficulties in planning and analyzing observations of the faint cluster
WDs.  While the main sequence typically stands out against the background
Galactic field stars, the WD sequence is often harder to distinguish.  The
contaminating objects in the WD region, however, are simulated Galactic
field WDs, whose number counts at these faint flux levels is unknown.  It
is also important to remove the abundant background galaxies at the faint
flux levels where the WDs are found, as these galaxies outnumber cluster
WDs and have similar colors.  These contaminating background galaxies are
not included in the cluster simulations, which assume some technique is
applied effectively to remove them.  The need to remove background
galaxies is one of the primary reasons HST is so appropriate for this
study.  Next generation ground-based telescopes with $I$-band AO systems
could also morphologically reject the background galaxies, but at present
this is not possible from the ground.  Among these simulated clusters some
(NGC 2243 and NGC 2660) display enough WDs that there is a clear pile-up
near the WD terminus.  Others (NGC 2204 and NGC 2506) should be uncrowded
enough that cluster WDs will dominate the lower left of the CMD.  The
remaining three clusters may pose greater difficulties, especially if
crowding is worse than predicted, though there are additional means to
extract the cluster WDs.  The easiest approaches are to observe additional
cluster fields in order to build up the number of cluster WDs, to observe
nearby control fields to better estimate Galactic field star and
background galaxy contamination, and to add a third, blue-sensitive filter
to the observation sequence.  While field WDs will not separate as well
from cluster WDs with the addition of another filter, the greater color
baseline and three filter information will help separate the warmer white
dwarfs (those with strong balmer lines, with $T_{\rm eff} \ga$ 8000 K)
from background main sequence stars.  Should these techniques prove
insufficient and where a particular cluster is a good example in age -
metallicity parameter space, a second HST epoch a few years later (King
\etal 1998) or ground-based epoch about a decade later (see Platais \etal
2003) can be obtained to isolate the cluster stars based on common proper
motion.

\section{Conclusion}

White dwarf cooling theory and very deep observations in star clusters
provide a new tool to test stellar evolution theory and time scales as
well as place two different age dating techniques on the same calibrated
scale.  Fortunately, directly comparing main sequence turn-off ages to
white dwarf ages is only weakly sensitive to realistic levels of errors in
cluster distance, metallicity, and reddening.  More generally, it is
encouraging to see the good overall agreement between WD and modern MSTO
ages over the range 0.1 to 4 Gyr.  Future application of WD isochrones to
open clusters with a variety of ages and metallicities, such as those open
clusters I have simulated, will test the consistency and limitations of
white dwarf and main sequence evolution theory.  Eventually, very deep
observations of globular clusters with HST and the ACS and future, large
ground-based facilities, calibrated by extensive HST and ground-based
observations and analyses of stars in open clusters, will yield accurate
and precise WD ages for the Galactic halo.  These same open cluster
observations will calibrate ongoing (Kilic \etal 2005) work on the age of
the Galactic disk via field white dwarfs.  This latter technique can date
individual WDs (Bergeron, Leggett, \& Ruiz 2001), and its improved
calibration will allow WD researchers to determine not just the age of the
Galactic disk, but also the age and age distribution of the Galactic thick
disk and halo.

\acknowledgments
I would like to thank Mukremin Kilic, Mike Montgomery, and Don Winget for
helpful discussions and the anonymous referee for a review that helped me
substantially improve this paper.  This material is based upon work
supported by the National Aeronautics and Space Administration under Grant
No.\ NAG5-13070 issued through the Office of Space Science, and by the
National Science Foundation through Grant AST-0307315.

\newpage

\figcaption{Main sequence turn-off versus white dwarf ages from recent
studies.  The WD age for the Hyades is a lower limit to a greater degree
than are the WD ages for the other clusters, since 50 to 90\% of the Hyades
has likely evaporated (Weidemann \etal 1992).}

\figcaption{The effect of three types of errors on the derived MSTO age
versus the ratio of the WD-to-MSTO age over the range of 100 Myr to 4 Gyr.
Calculations are performed at log(age) = 8.0, 8.3, 8.6, 9.0, 9.3, and 9.6.
The thick lines show the effect of overestimating distance by 0.2 mag.
The thin lines show the effect of overestimating metallicity by 0.2 dex.
The dotted lines show the effect of overestimating metallicity by 0.2
dex, then compensating by decreasing reddening to keep the MSTO at the
same color.}

\figcaption{The three categories of error presented in the MSTO age versus
WD age diagram.  The symbols are the same as in Figure 2.  The diagonal
dotted line crossing most of the diagram is the one-to-one correspondence
line where MSTO ages and WD ages are identical.}

\figcaption{Simulated $VI$ color magnitude diagrams for the open cluster NGC 1245, 
with cluster parameters as listed in Table 2.}

\figcaption{Same as Figure 4, but for NGC 2204.}

\figcaption{Same as Figure 4, but for NGC 2243.}

\figcaption{Same as Figure 4, but for NGC 2360.}

\figcaption{Same as Figure 4, but for NGC 2506.}

\figcaption{Same as Figure 4, but for NGC 2660.}

\figcaption{Same as Figure 4, but for NGC 7789.}

\newpage

\begin{figure}[!t]
\plotone{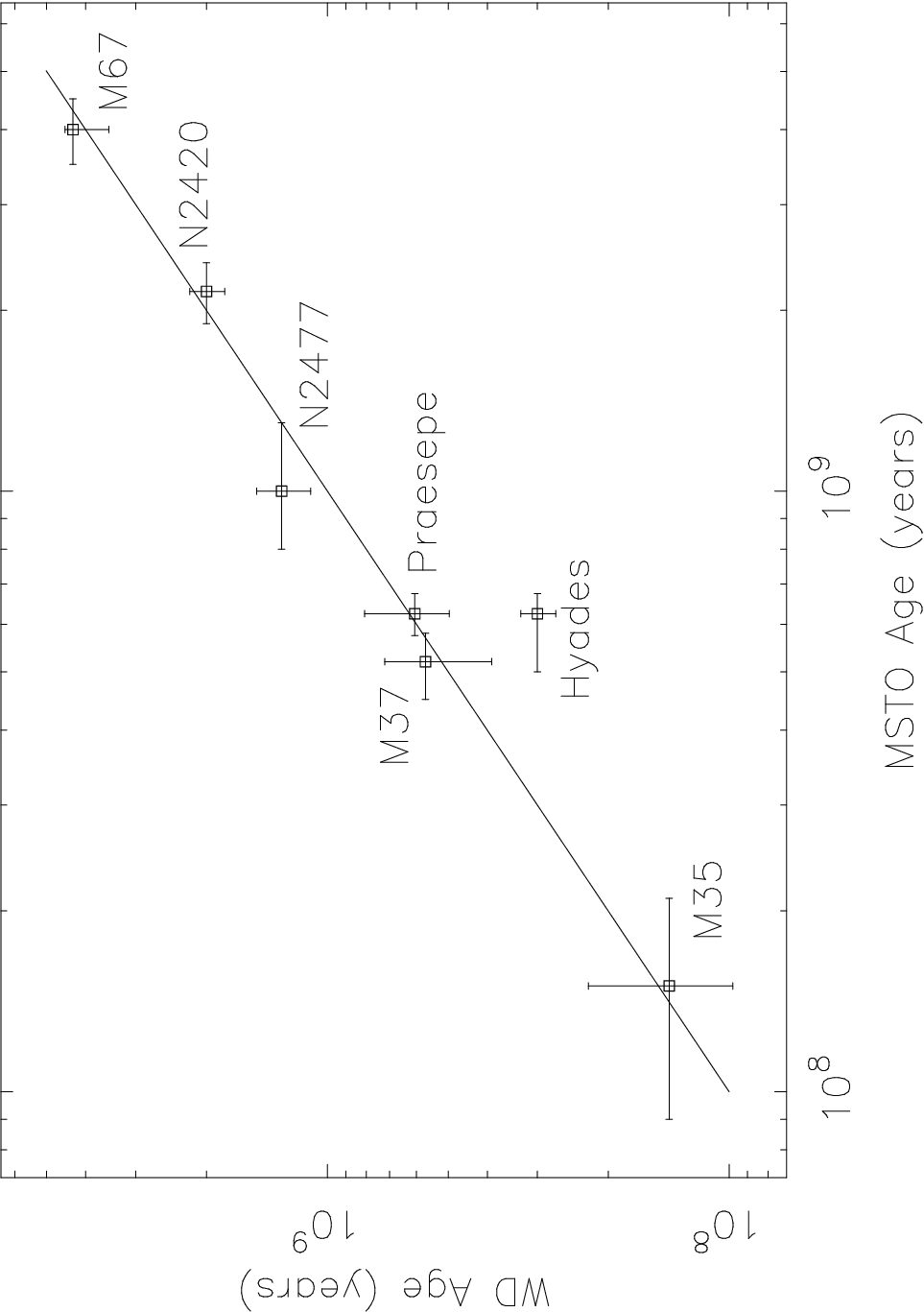}
\end{figure}

\begin{figure}[!t]
\plotone{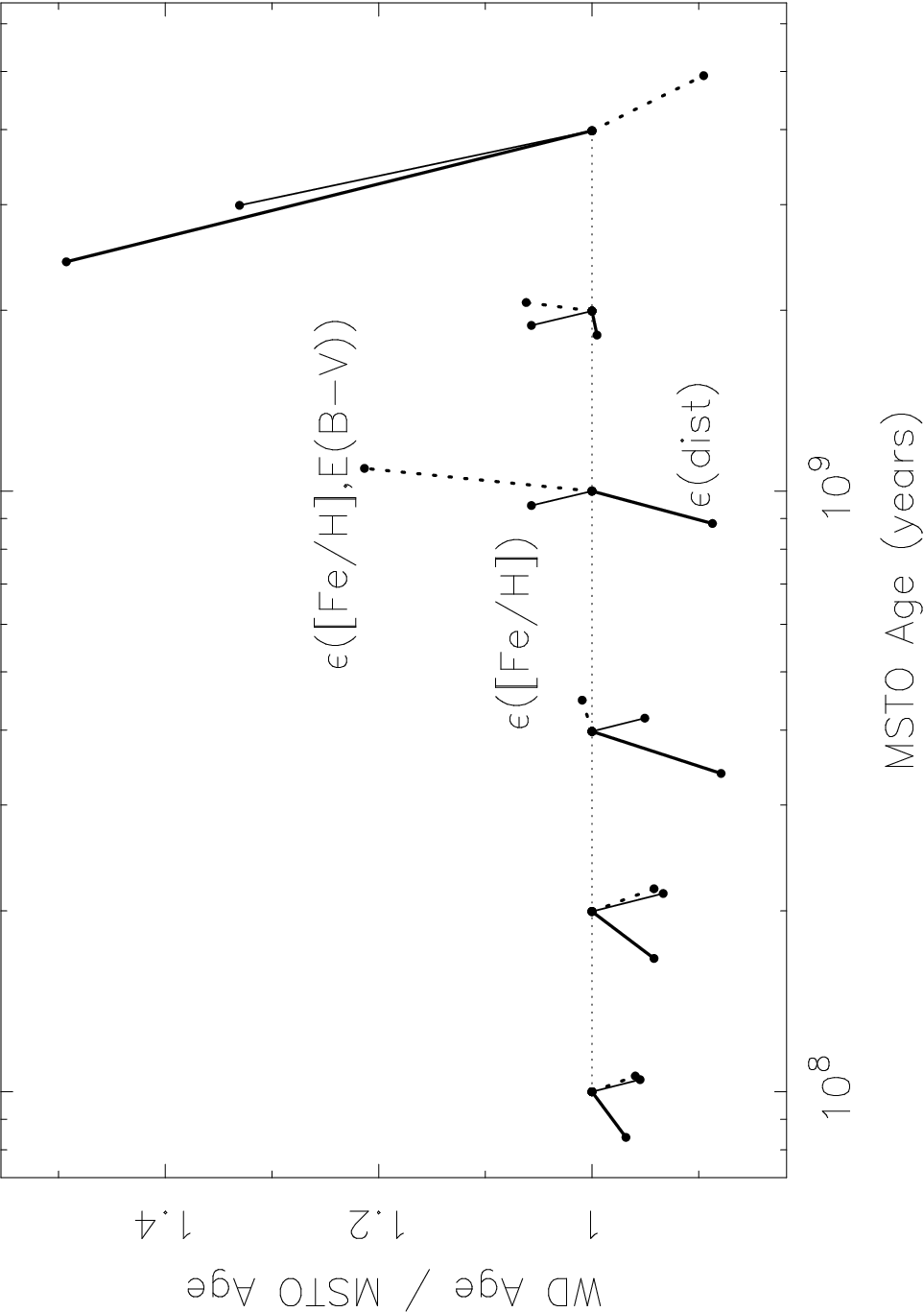}
\end{figure}

\begin{figure}[!t]
\plotone{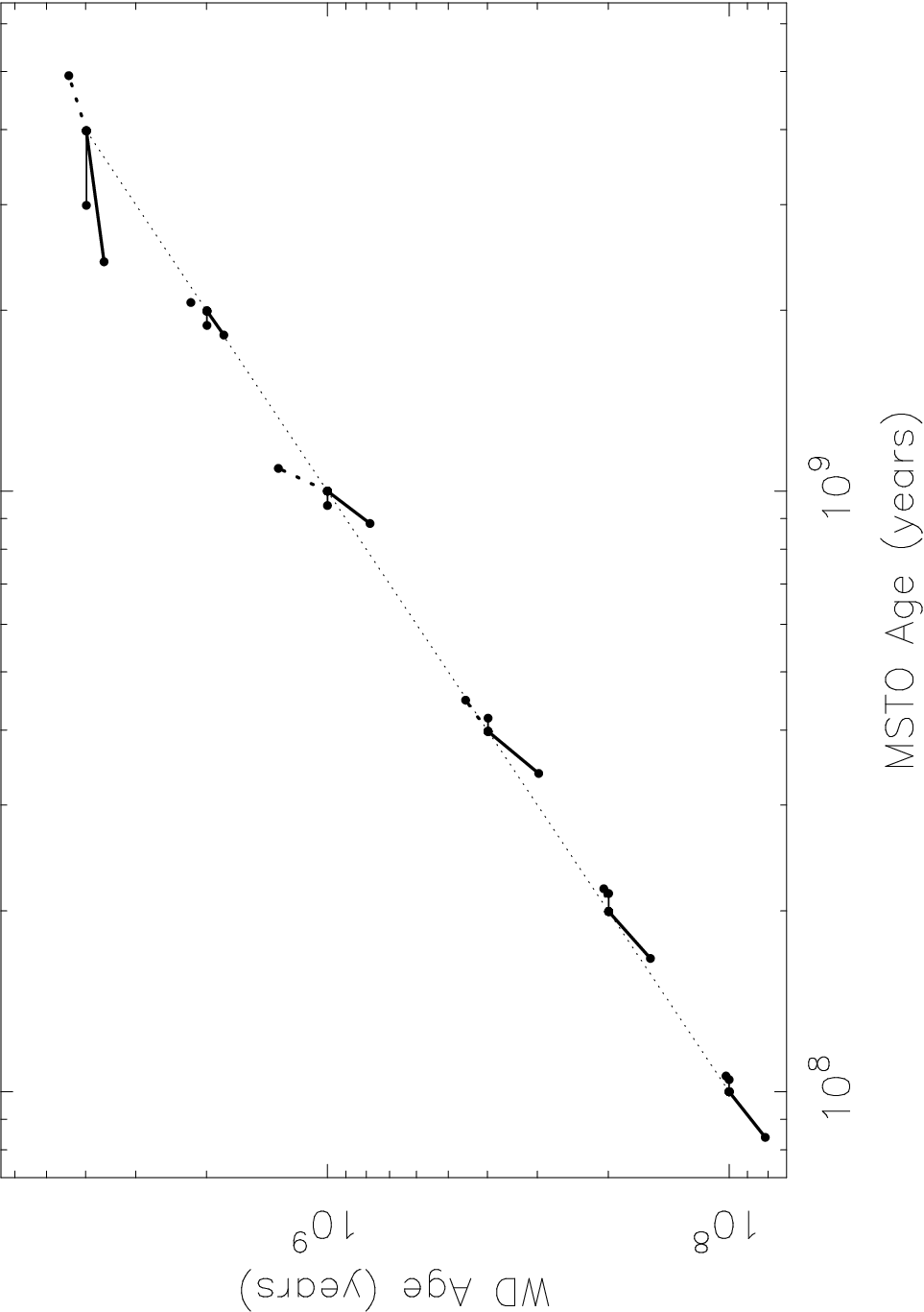}
\end{figure}

\begin{figure}[!t]
\plotone{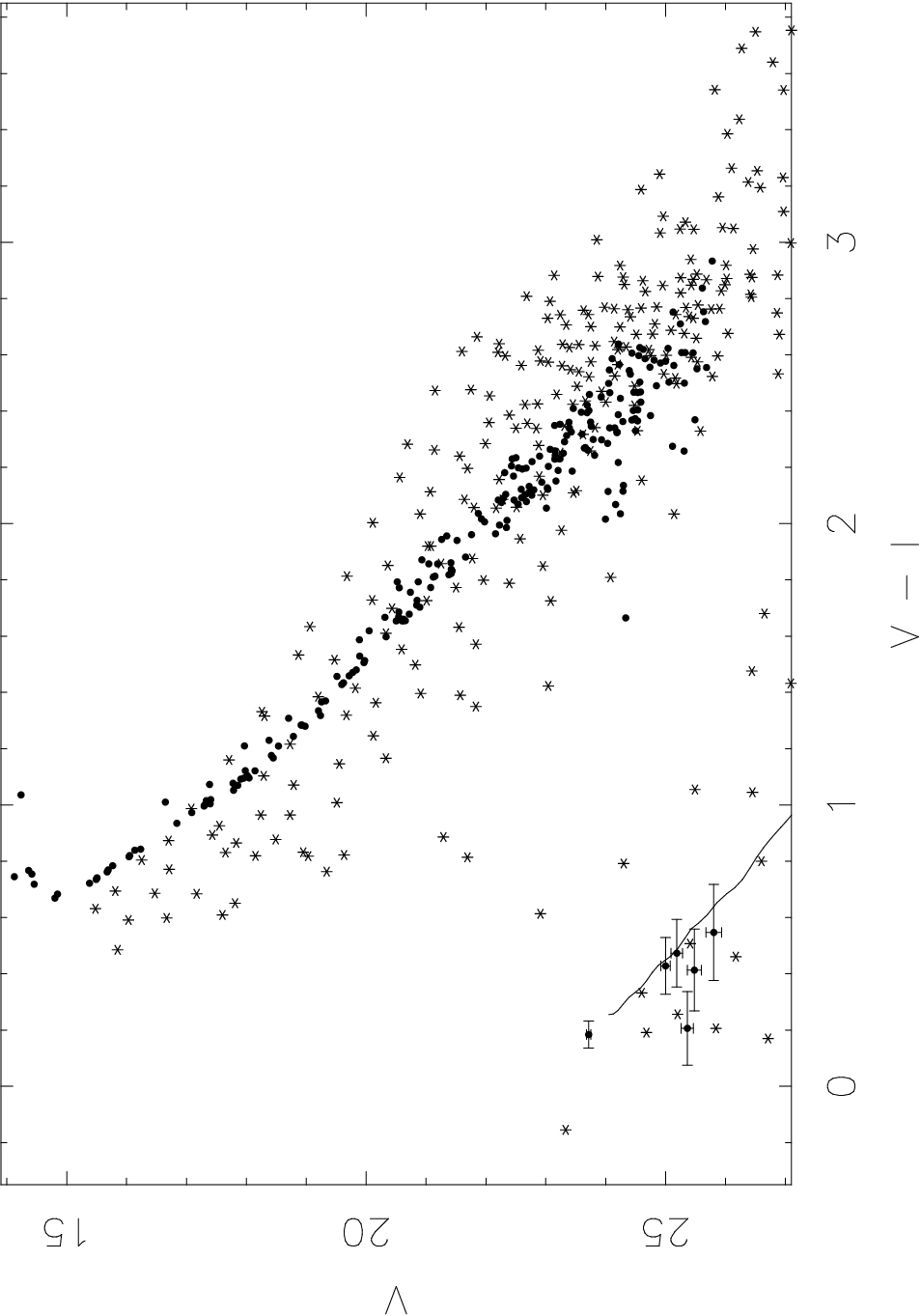}
\end{figure}

\begin{figure}[!t]
\plotone{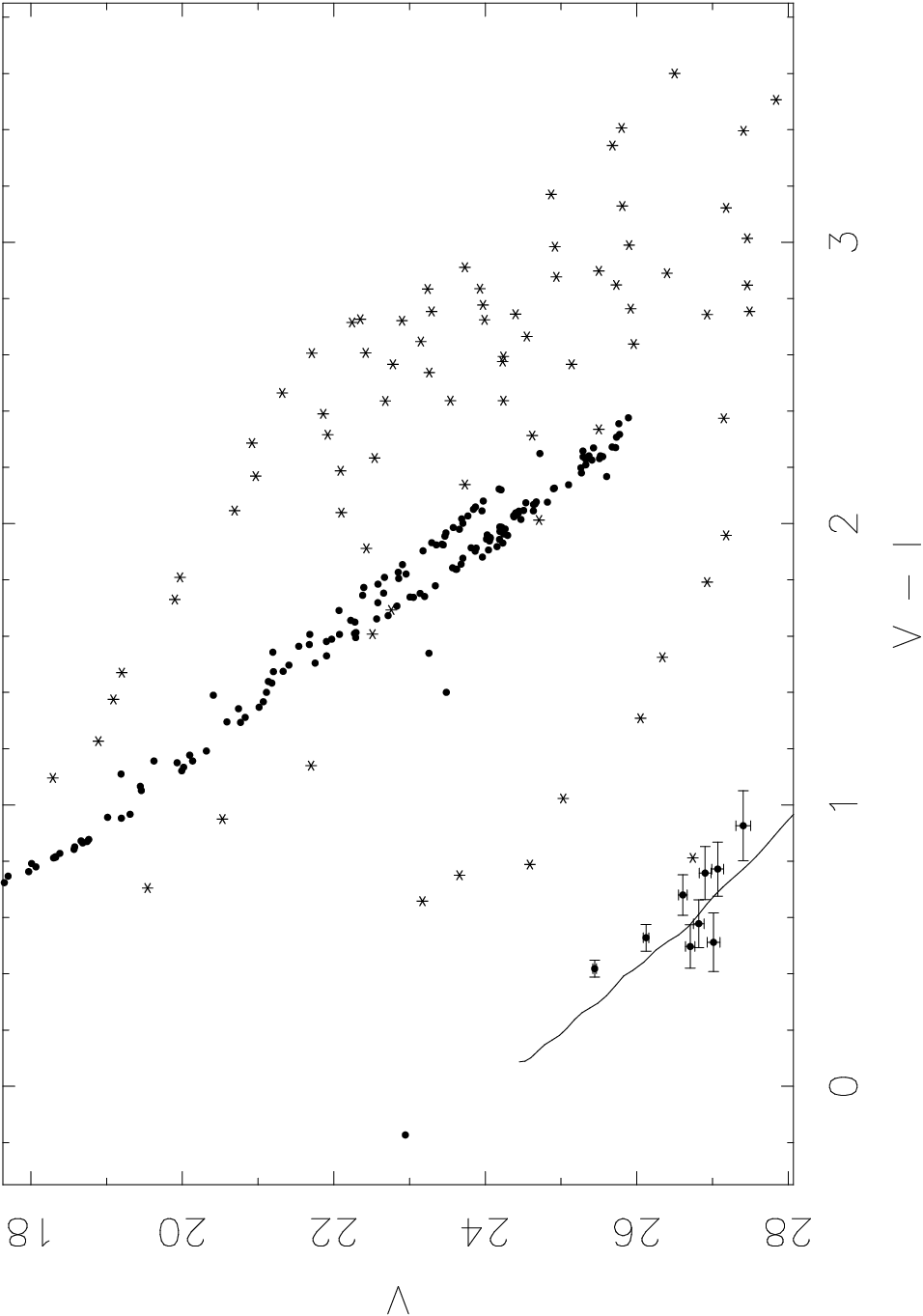}
\end{figure}

\begin{figure}[!t]
\plotone{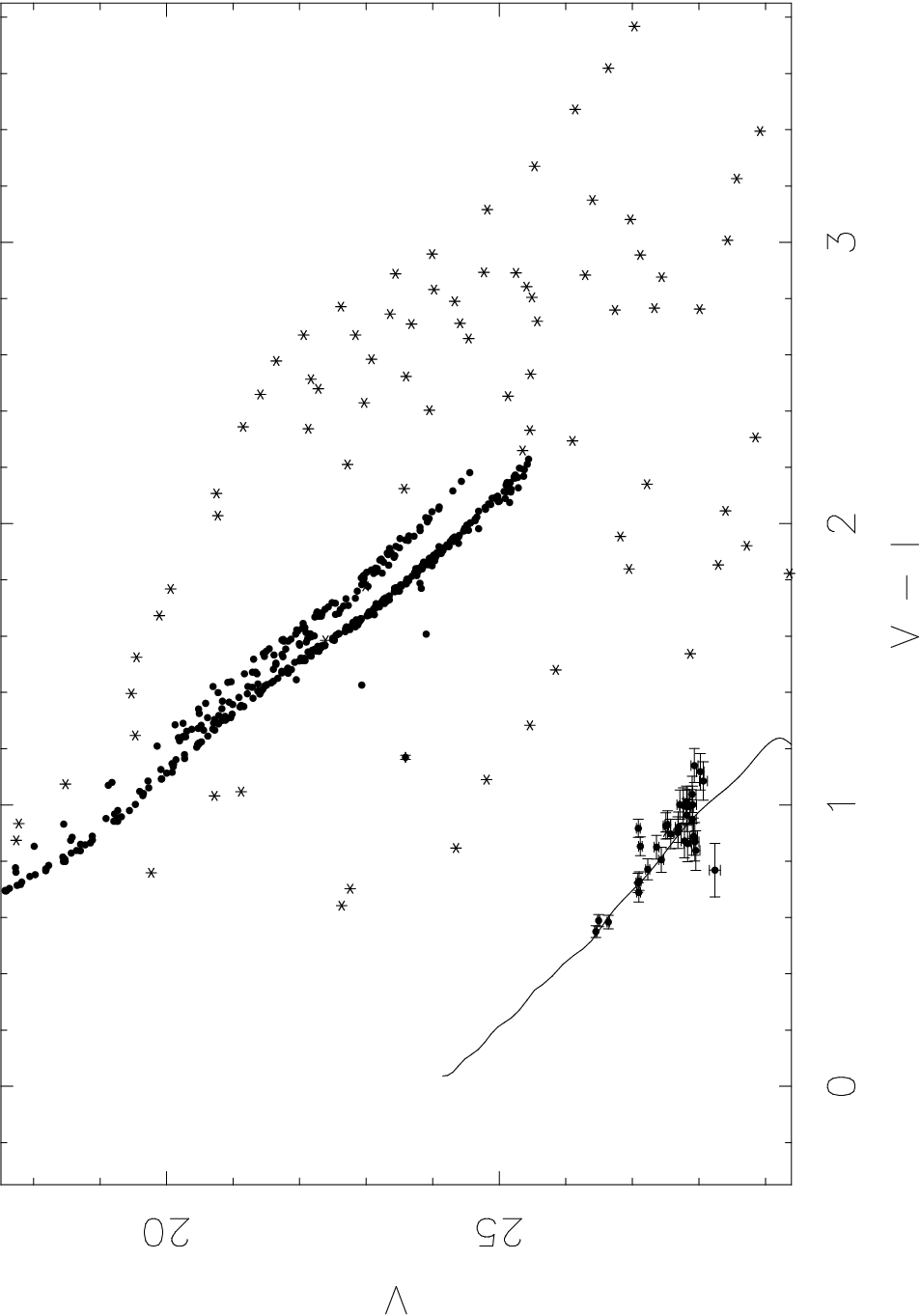}
\end{figure}

\begin{figure}[!t]
\plotone{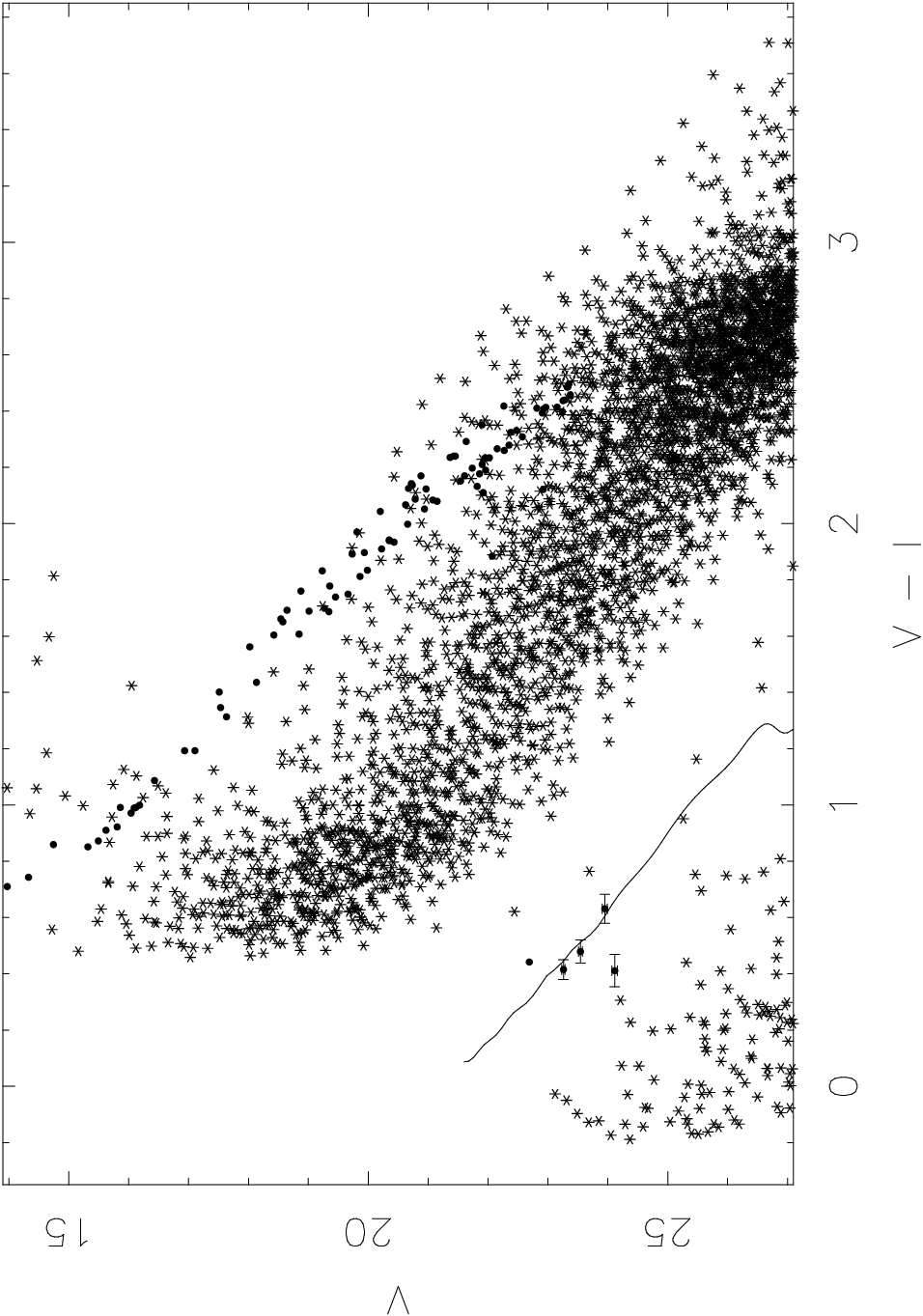}
\end{figure}

\begin{figure}[!t]
\plotone{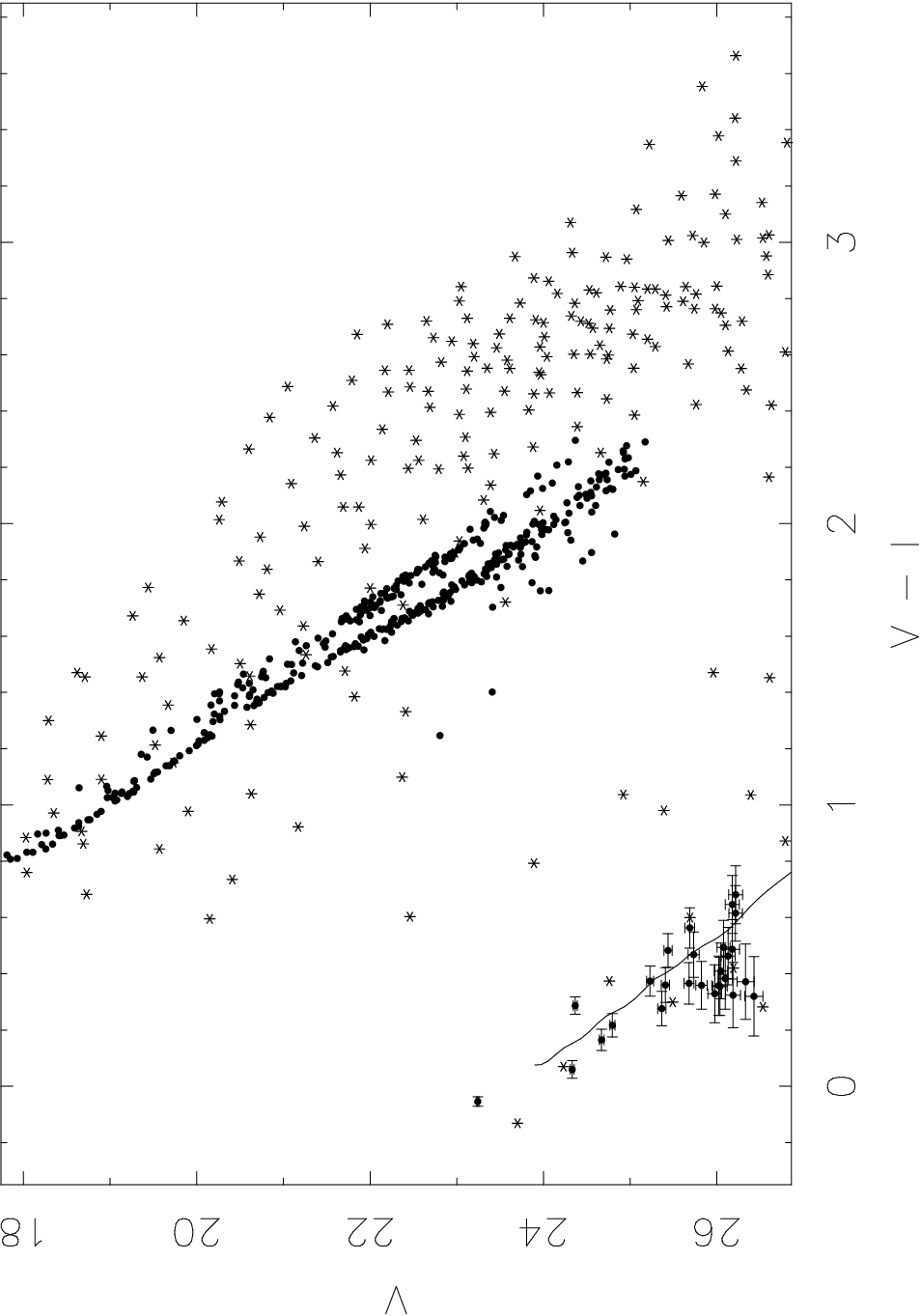}
\end{figure}

\begin{figure}[!t]
\plotone{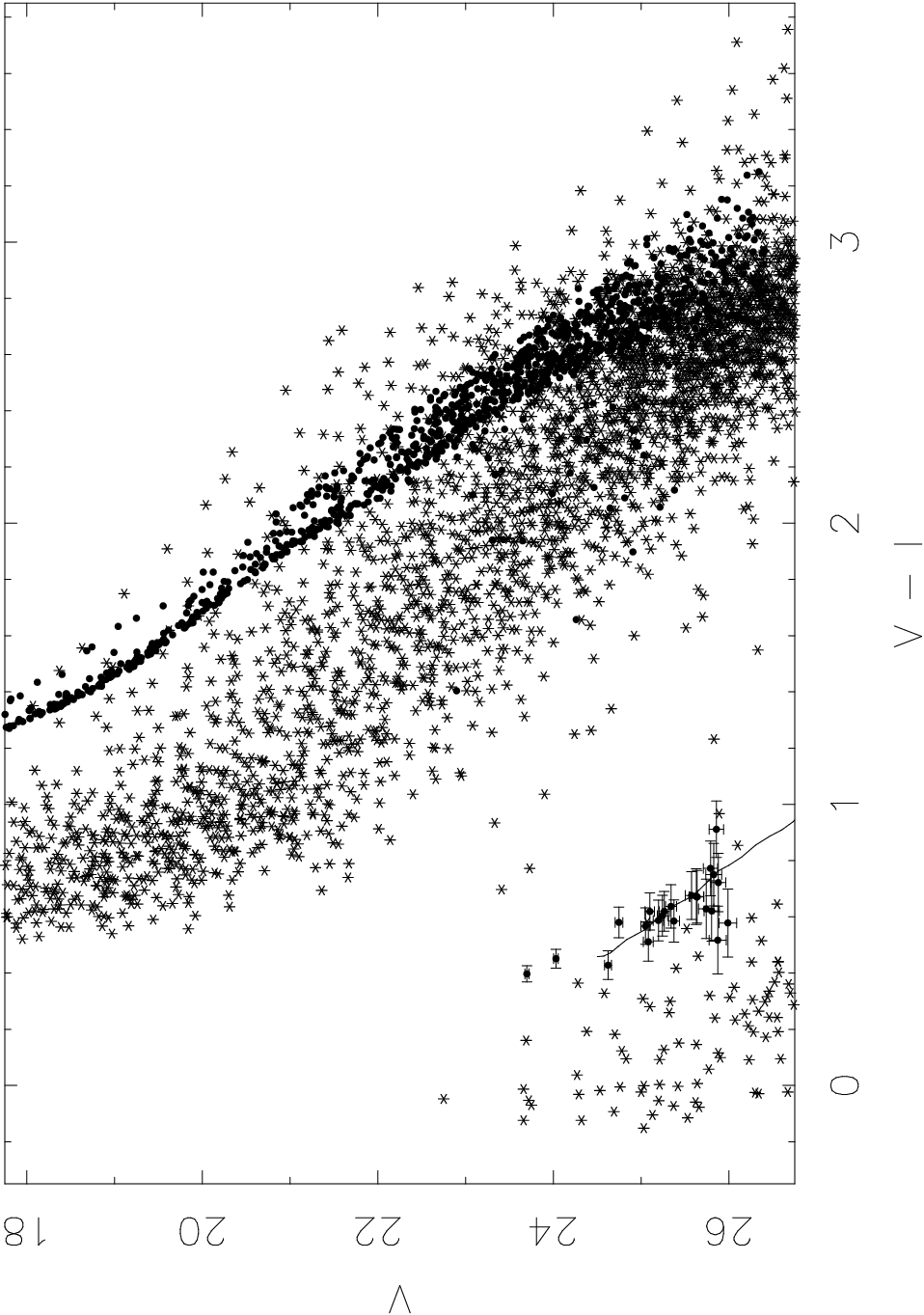}
\end{figure}

\begin{figure}[!t]
\plotone{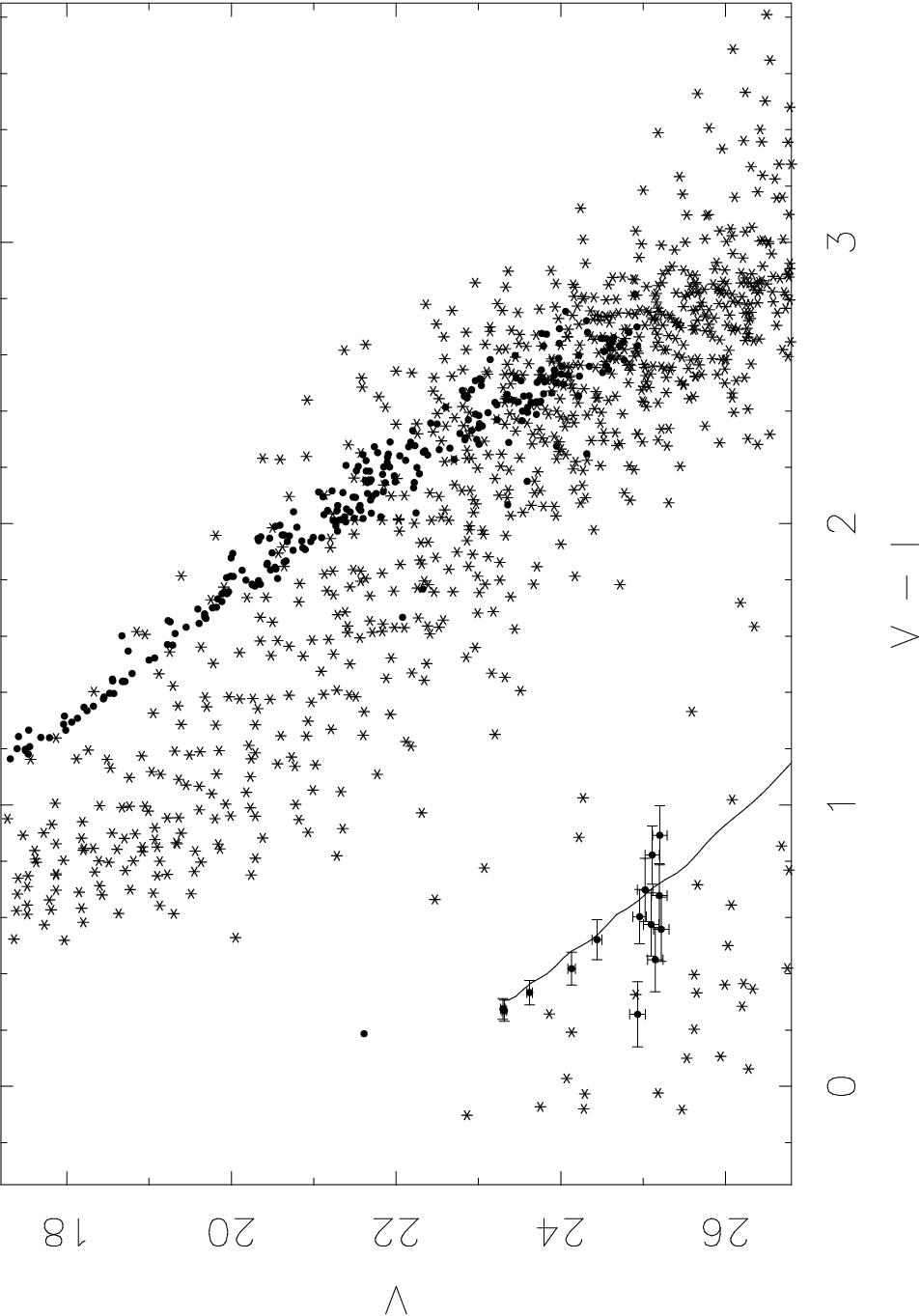}
\end{figure}

\clearpage

\begin{deluxetable}{llll}
\tablewidth{0pt}
\tablecaption{White Dwarf Versus Main Sequence Ages for Open Clusters}
\tablehead{
\colhead{Cluster} & \colhead{WD Age (Gyr)} & 
\colhead{MSTO Age (Gyr)} & \colhead{Ref} \\
\phantom{12} (1) & \phantom{1234} (2) & \phantom{1234} (3) & (4)}
\startdata
M 35     & 0.141 (+0.083,$-$0.043) & 0.150 ($\pm$0.06)      & 1 \\
Hyades   & 0.3   ($\pm$0.03)       & 0.625 (+0.05,$-$0.125) & 2 \\
M 37     & 0.57  (+0.15,$-$0.18)   & 0.52  (+0.6,$-$0.7)    & 3 \\
Praesepe & 0.606 (+0.202,$-$0.109) & 0.625 ($\pm$0.05)      & 4 \\
NGC 2477 & 1.3   ($\pm$0.2)        & 1.0   (+0.3,$-$0.2)    & 5 \\
NGC 2420 & 2.0   ($\pm$0.2)        & 2.15  ($\pm$0.25)      & 6 \\
M 67     & 4.3   (+0.2,$-$0.8)     & 4.0   ($\pm$0.5)       & 7 \\
\enddata
\tablerefs{
1. The WD age is derived from the cooling age for the oldest cluster WD
from Williams, Bolte, \& Koester (2004) plus a precursor age of $\sim56$
Myr, based on the object's initial mass ($\sim7 M_\sun$), which was
calculated from the Weidemann (2000) initial-final mass relation.  The
MSTO ages are based on Grocholski \& Sarajedini (2003), Steinhauer (2003),
and Steinhauer \& Deliyannis (2004).
2. The WD age is from Weidemann \etal (1992) and the MSTO age is from
Perryman \etal (1998).
3. The WD and MSTO ages are from Kalirai \etal (2001).  The MSTO age range
is extracted from their discussion as it is not explicitly presented.
4. The WD age is derived from the cooling age of the oldest WD from Dobbie
\etal (2005) plus a precursor age of 106 (+109,$-$43) Myr based on the
object's initial mass of 5.3 (+1.4,$-$1.3) $M_\sun$, in turn derived from
the Weidemann (2000) initial-final mass relation.  The MSTO age is from
Perryman \etal (1998).
5. The WD age is from von Hippel, Gilmore \& Jones (1995) and the MSTO age
is from Kassis \etal (1997).
6. The WD age is from von Hippel \& Gilmore (2000) and the MSTO ages
represent the mean and range of the convective overshoot results of
Carraro \& Chiosi (1994); Demarque, Sarajedini \& Guo (1994); Lee \etal
(1999); Pols \etal (1998); and Twarog, Anthony-Twarog \& Bricker (1999).
7. The WD age is from Richer \etal (1998) and the MSTO ages are from
Demarque, Green \& Gunther (1992) and Dinescu \etal (1995).
}
\end{deluxetable}

\begin{deluxetable}{llllrrc}
\tablewidth{0pt}
\tablecaption{Open Cluster Parameters for Simulations}
\tablehead{
\colhead{Cluster} & \colhead{$V-$M$_{\rm V}$} & \colhead{A$_{\rm V}$} & 
\colhead{Age} & \colhead{\it l} & \colhead{\it b} & \colhead{Ref} \\
\phantom{123} (1) & \phantom{1} (2) & (3) & (4) & (5) \phantom{1} & (6) \phantom{1} & (7)}
\startdata
NGC 1245 & 12.95 & 0.68 & 1.0 & 146.63 &  $-$8.92 & 1 \\
NGC 2204 & 13.35 & 0.25 & 2.5 & 226.02 & $-$16.07 & 2 \\
NGC 2243 & 13.05 & 0.12 & 5.  & 239.50 & $-$17.98 & 3 \\
NGC 2360 & 10.5  & 0.25 & 2.2 & 229.80 &  $-$1.40 & 4 \\
NGC 2506 & 12.8  & 0.22 & 1.9 & 230.60 &     9.97 & 5 \\
NGC 2660 & 13.4  & 1.2  & 0.9 & 265.84 &  $-$3.03 & 6 \\
NGC 7789 & 12.2  & 0.81 & 1.6 & 115.49 &  $-$5.35 & 7 \\
%
%
\enddata
\tablerefs{
1. Burke \etal 2004
2. Frogel \& Twarog 1983
3. Bergbusch, Vandenberg, \& Infante 1991
4. Mermilliod \& Mayor 1990
5. Marconi \etal 1997
6. Sandrelli \etal 1999
7. Gim \etal 1998
}
\end{deluxetable}

\end{document}